# CVA6's Data cache: Structure and Behavior

Valentin Martinoli[1,2], Yannick Teglia[1], Bouagoun Abdellah[1], Régis Leveugle[2]
[1] Cybersecurity Hardware lab, Thales DIS, La Ciotat, France, valentin.martinoli@external.thalesgroup.com, yannick.teglia@thalesgroup.com, abdellahbouagoun@gmail.com
[2] Univ. Grenoble Alpes, CNRS, Grenoble INP*, TIMA, 38000 Grenoble, France, firstname.name@univ-grenoble-alpes.fr
*Institute of Engineering Univ. Grenoble Alpes

## Abstract

Since Spectre [1] and Meltdown's [2] disclosure in 2018, a new category of attacks has been identified and characterized by the scientific community. These attacks, called micro architectural attacks, take advantage of optimization mechanisms implemented inside deep architectural elements to induce a leakage of secret material like encryption keys. The Foreshadow attack [3], which was the first one to target Intel's secure enclave technology (namely SGX [4]) has been developed shortly after. It opened the way to micro architectural attacks on Intel's architecture, and led to the quick development of micro architectural attacks until today. While Spectre and Meltdown are often considered as the first micro architectural attacks, one can argue that cache attacks, as introduced by Osvik et al. in 2006 [5], can be seen as the first types of micro architectural attacks that were developed. Now, even though there are many variants [6], they are still the most prominent type of micro architectural attacks. One example of cache micro architectural covert-channel is the Prime+Probe [7].

Lately targeting the Intel architecture, the micro architectural attacks are now challenging a wider variety of CPUs. Recently, CPUs running the RISC-V Instruction Set Architecture [8] have been targeted. One famous and widely used RISC-V CPU is the ETH Zurich's CVA6 [9] [10] (formerly Ariane) core. CVA6 is a 6-stage, single issue, in-order CPU.

To the best of our knowledge, there is no existing document presenting very detailed aspects of the CVA6's micro architecture, especially with respect to the data cache. Such information is mandatory to deeply understand any architectural or micro architectural study successfully, such as the replication of the Prime+Probe attack on the CVA6 CPU proposed by Nils Wistoff.

This paper presents the implementation of the Data cache in the CVA6 CPU from OpenHW Group by focusing on its memory structure and explaining through several examples what happens when a request for memory allocation occurs.

# Contents

# 1. The CVA6 Core

The CVA6 core has been developed by ETH Zürich and University of Bologna. It is an open-source application-class 64-bit processor, implementing the RISC-V Instruction Set Architecture. It has been developed in System Verilog, and is available online[1]. This core is Linux-capable and can run the M, S and U privilege modes. It contains a Translation Lookaside Buffer (TLB), tightly integrated Data and Instruction caches and a hardware Page Table Walker (PTW). CVA6 has been optimized for performance. With GlobalFoundries 22FDX technology, the core achieves a frequency up to 1.7GHz and up to 40Gop/sW peak efficiency. CVA6's pipeline (shown in Figure 1) is made of six stages. The issue and commit of instructions are both in-order. It is worth noting that the CVA6 core is not fully in order. Indeed, the execution stage is partly out-of-order, specifically, the write back happens out of order. The scoreboard structure serves as a reordering buffer for the in-order architectural commit to work properly. This particularity has several effects on the core's structure. The CVA6 also contains a branch-prediction unit.

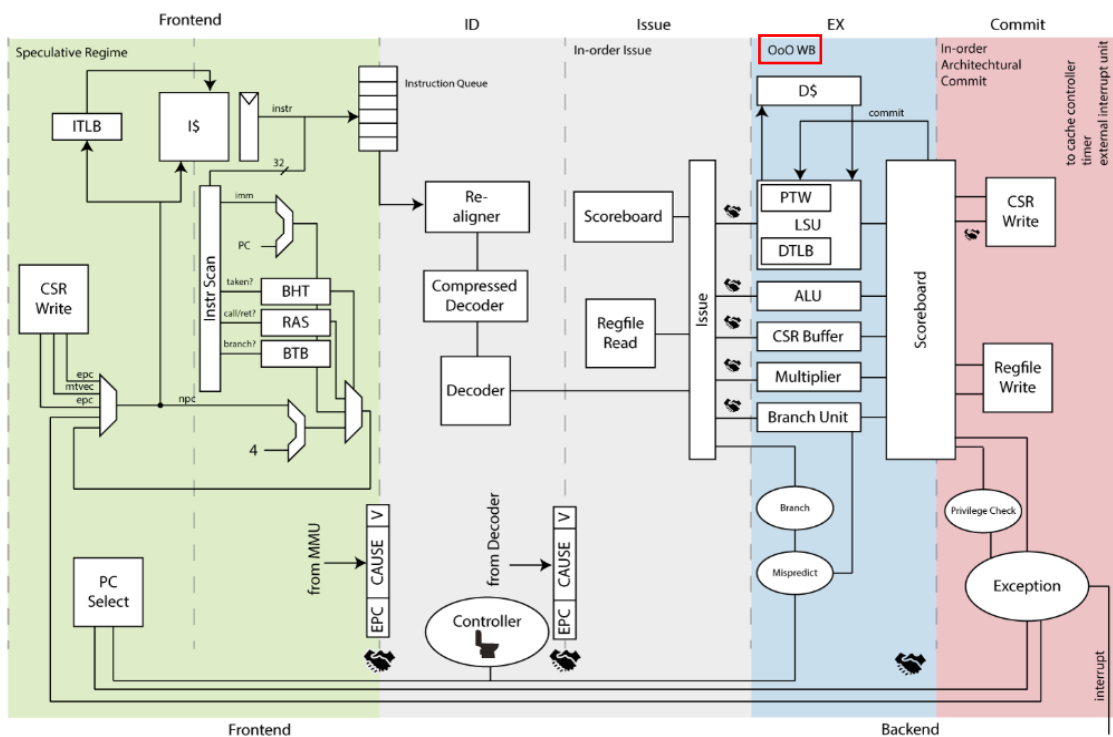

*Figure 1 – CVA6's Pipeline [9]*

The Data cache inside the CVA6 CPU is a complex subsystem, composed of several functional blocks interacting continuously with each other. The Figure 2 represents the whole Data cache's structure and the components it cooperates with directly. The Data cache mainly interacts with the CPU, and the main memory.

The study presented in this document focuses on one of the configurations of the core as available by default directly from the Open Hardware group[1].

---

[1] https://github.com/openhwgroup/cva6

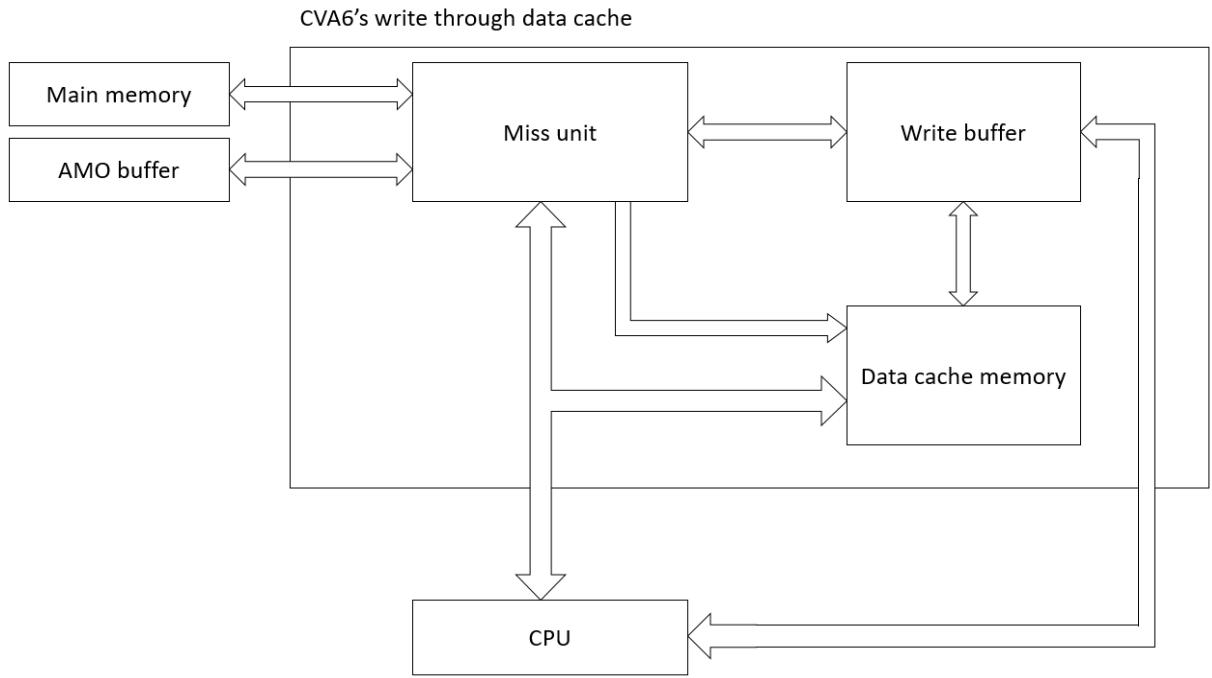

*Figure 2 - Representation of the data cache in the CVA6 core*

## 2. CVA6's data cache: characteristics and general behavior

### 2.1. Dimensioning and behavior

The default cache in the studied configuration is a 32 KB 8-way set-associative data cache with 16-byte line sizes. It is configured in a write through no write-allocate configuration that will be described later in the document. The different other configuration possibilities will be mentioned and explained but will not be detailed further.

The data cache is composed of 256 sets. Each set is composed of eight cache lines called "ways". For the rest of the document the term "way" will therefore be used. Each way contains 16 bytes according to the standard configuration of the data cache. Based on this information, one can conclude that the size of the L1 data cache is 256 (sets) × 8 (ways) × 16 (bytes in one way) = 32KB. The addresses in the CVA6 CPU are 64-bit long, and the data handled is the size of a single way (i.e. 32 bytes). The cache addresses are physical (and not virtual). Therefore, the addresses' sizes in the data cache vary with the cache's size. The data and instruction caches are both linked to an AXI adapter module in order to connect them to a 64-bit AXI bus.

### 2.2. How addresses can be decomposed to represent the CVA6's data cache

Considering the previously presented architecture, it is now possible to specify how to decompose a specific address in order to find which area in the cache it represents. For the CVA6's data cache, the physical addresses are 64-bit long and are composed of the following fields:

- 0 to 3: Offset => it specifies the location of a byte inside the way and thus enables to choose the appropriate data bank
- 4 to 11: Index => these bits allow to distinguish between the 256 available Sets
- 12 to 63: Tag => these 52 bits are used to represent the requested memory location

For clarification purposes, an example will now be detailed in order to put into practice the previously presented decomposition. In this example, the following address will be considered: 0x0000008000b010. By applying the decomposition previously introduced, we obtain the following result as shown on Figure 3:

|  | Dcache address | | |
|---|---|---|---|
|  | Tag | Index | Offset |
| Hexadecimal | 000000008000b (52bits) | 01 | 0 |
| Binary |  | 00000001 | 0000 |

*Figure 3 - Decomposition of the example address 0x000000008000b010*

### 2.3. CVA6's writing/eviction policies

When considering a cache, it is mandatory to look at the different policies and configuration used inside. A cache hit consists in requesting a data that is already stored in a memory block inside the Data cache. A cache miss is a request to a data that does not have an attributed memory block inside the Data cache.

There are two main possible writing policies:

- The CPU updates the cache and the main memory synchronously when there is a request to modify the content of the accessed memory block. This is a **Write through** policy.
- The CPU only updates the cache; it writes the data to the main memory only when there is another request to overwrite an attributed cache line that is "dirty", i.e. where values have been modified by the processor. This is done to avoid losing the content previously stored in the Data cache, while avoiding too many main memory accesses that are useless for temporary data quickly changed by the processor. It is called a **Write back** policy.

Moreover, there are also two possible approaches to handle write operations when the data is not already allocated in the main memory:

- **Write-allocate** consists in writing the data in the cache and then into the main memory
- **No-write-allocate** consists in writing the memory block only in the main memory and not in the cache. Data is only loaded into the cache on read misses.

The default configuration of the CVA6's data cache that is considered in this document is a **write through, no-write-allocate** data cache. This means that upon writing a new data that is not already allocated inside the cache, it will first be stored only in the main memory. Then on the next read of this data, it will be uploaded from the main memory and written inside the data cache.

The eviction policy is necessary in case all the potential cache lines for a required data are already filled and "valid". In that case, a choice must be made about which line will be evicted. In the CVA6 core the index of the cache line to be evicted is the output on a LFSR (Linear-Feedback Shift Register). The CVA6, in the default configuration considered here, uses a parametric LFSR with precomputed coefficients for LFSR lengths from 4 to 64 bits. However, only the 3 lowest bits are used. This depends on the cache's parameters that can be changed by the developer. This LFSR serves as a pseudo-random number generator and provides a pseudo random data when a cache line has to be evicted. Upon a cache miss, and if all the cache lines are valid, then the LFSR's output obtained is used as the index of the cache line to be evicted. Otherwise, if there are some empty cache lines (i.e. with their valid bit equal to 0), the first line with a non-valid bit (i.e. the line with the lowest index having its validity bit set to 0) is selected and used to store the new data. Therefore, the CVA6 core uses a pseudo-random eviction policy when the cache is filled. The LFSR uses Galois LFSR feedback masks that were taken online[2]. These coefficients are hard coded in the sources and a XOR operation is applied between the input of the LFSR (i.e. the output of the previous round) and these masks. Moreover, additional block cipher layers can be instantiated to non-linearly transform the pseudo-random LFSR sequence at the output, and hence break the shifting patterns. The RTL contains the code for these cipher layers that can be used to transform the original linear transformation into a non-linear one that is therefore less predictable. The additional cipher layers can only be used for an LFSR width of 64 bits, since the block cipher has been designed for that block length. All of this eviction process is handled by the "miss unit" component, which will be detailed further in the next part.

## 3. The data cache memory's structure

The main component inside the Data cache's structure is the Data cache Memory. It contains all the data stored inside the Data cache at a given time. It is framed in orange on the Figure 4.

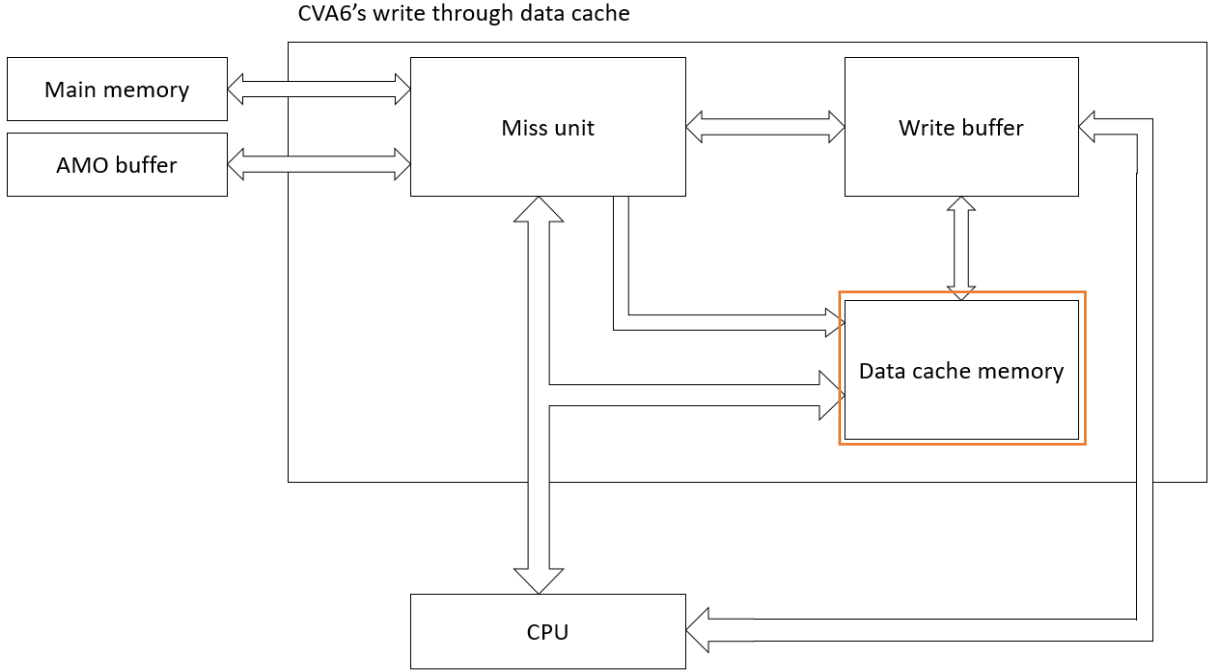

*Figure 4 - Localization of the Data cache Memory inside the data cache's structure of the CVA6 core*

The Data cache Memory is composed of two main structures: the data array, and the tag array. The former is used to store the raw data while the latter stores a part of the address used to access a specific region inside the Data cache. The next sections will successively present the structures of the data array and of the tag array.

---

[2] https://users.ece.cmu.edu/~koopman/lfsr/

### 3.1. How data is stored

The Data array is used to store the data inside the Data cache. It has a specific organization and structure that will now be developed.

The CVA6's data array consists of 256 Sets, indexed from 0 to 255 as shown in orange on the Figure 5. As for the specific implementation of this structure, two SRAM matrices, called banks are used. They are represented in green on the Figure 5. Each bank is a block of 256 entries (half sets, composed of 8 half ways) and each half set contains 8×64 bits (eight half ways). Using this structure, it is possible to dedicate bank 0 to storing the 64 Least Significant Bits (LSB) of each way (represented as the w0 part of each way on Figure 5) for the 256 sets, and bank one for the 64 Most Significant Bits (MSB) of all the ways (represented as the w1 part of each WAY on Figure 5).

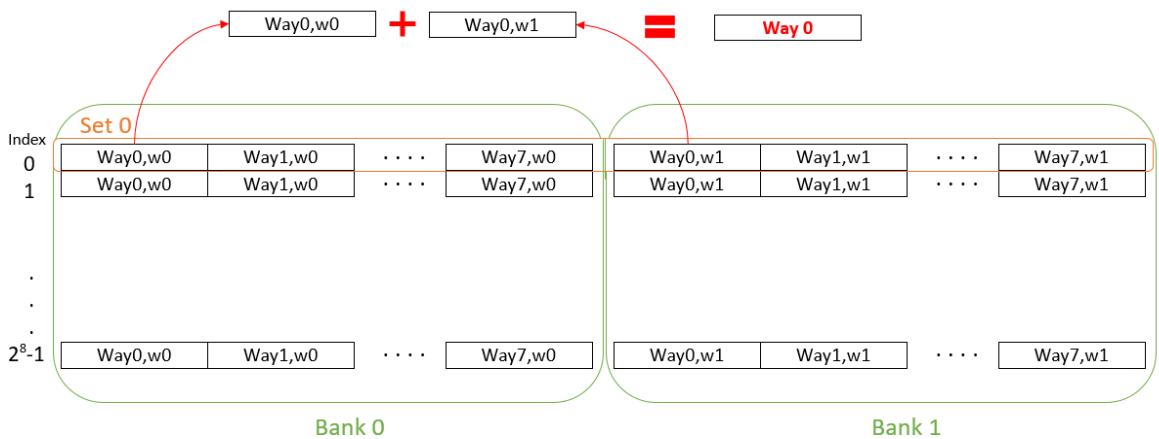

*Figure 5 - Implementation of the Data array in the L1 Data cache of CVA6*

### 3.2. How the cache accesses a specific data using the previous decomposition

A given address enables to access a cache set. However, a single set may contain several different memory locations. Therefore, tags are required to specify the rest of the address bits in order to be able to distinguish between the different memory locations mapping to the same set. Let us consider two data cache addresses: 0x0000008000b010 and 0x00000081234010. These two addresses map to the set number 1 in the bank 0. If we remove the information carried by the tag, this leads to considering both addresses as 0x010 (only considering the index and the offset). These two addresses map to the same set in the same bank as their indexes and offsets are equal. In such a situation, the tag is required to specify exactly which way is considered and to enable addressing at the way granularity level.

In the CVA6 core, the tag array is implemented using eight SRAM matrices, as represented on the Figure 6. Each matrix, numbered from zero to seven, is composed of 256 slots of 53 bits: 52 bits for the tag itself and one validity bit. Therefore, the SRAM matrix number "n" will contain the tags dedicated to the WAY number "n". For example, the tag for the $5^{th}$ WAY of the set number 125 will be the entry number 125 inside the $5^{th}$ SRAM matrix.

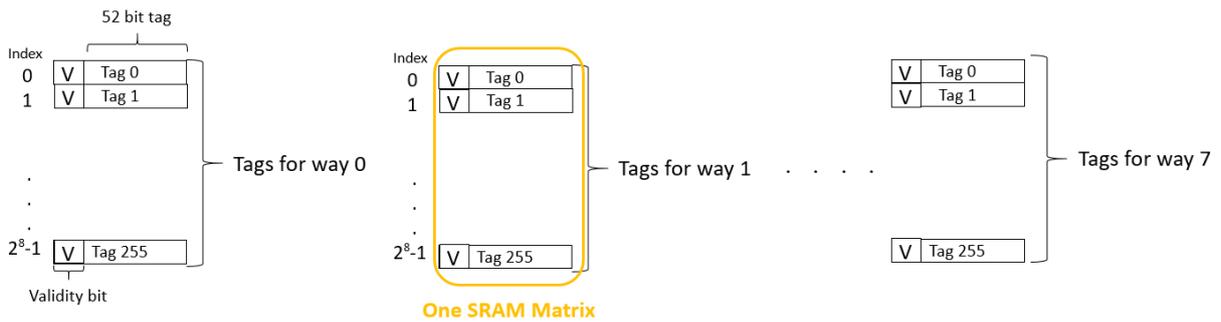

*Figure 6 - Implementation of the Tag array in the L1 Data cache of CVA6*

For clarification purposes, we will now reuse the previous addressing example in order to put into practice the previously presented structures and elements. In this example, the address considered was 0x0000008000b010. The decomposition previously introduced is depicted on Figure 3. Here, the offset is equal to 0x0, or 0000 in binary.

The MSB of the offset is used to choose which bank to target in the Data array:

- For offsets between zero and seven, the MSB is equal to zero, which means that the bank 0 will be selected.
- For the remaining values (8 to 15), the MSB is equal to one, meaning that the bank 1 will be selected.

For the address 0x0000008000b010, the bank number zero will be used.

The index's value is 0x01, or 00000001. This index refers to the second set in the data array. This leads to the selection of the eight tags that refer to the second set, with an index equal to one. The selection for this example is represented on the Figure 7.

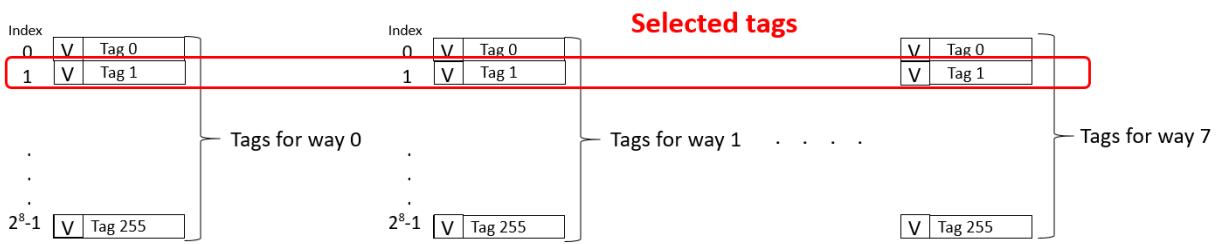

*Figure 7 - Selection of the tags corresponding to the second set (e.g. index equal to one)*

The tag is equal to 0x0000008000b. Among the previously selected tags, the number of the SRAM matrix of the corresponding entry will give out the way targeted by the address considered. The entry from the eighth SRAM matrix is also equal to 0x0000008000b, meaning that the tag refers to the way number seven.

As a conclusion, the requested 64-bit memory block with the address 0x0000008000b010 is in bank 0, 2$^{nd}$ Set (i.e., the set with the index equal to 1), 8$^{th}$ WAY. The results developed above are represented on the Figure 8.

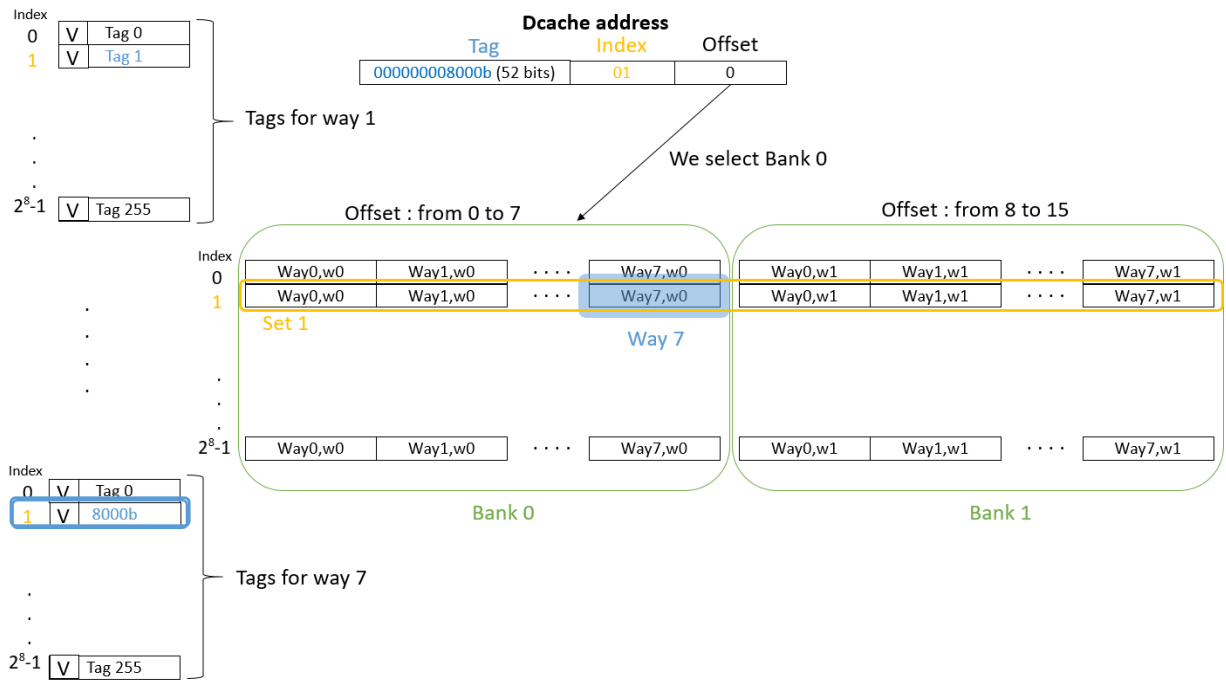

*Figure 8 - Request of a 64-bit memory block with the address 0x0000008000b010 from the Data cache of the CVA6 core*

## 4. Other components working with the data cache's memory

### 4.1. Storing atomic operations in the AMO buffer

The AMO buffer is located outside of the Data cache's structure. It interacts with the Miss unit and is used to buffer an atomic memory operation for the cache subsystem. It also handles interfacing with the commit stage, acting as an intermediate between the Data cache subsystem and the pipeline's commit stage.

The AMO buffer issues requests to the cache subsystem by providing a physical address and a data for the requested store in the Data cache. The information contained in these requests comes from the commit stage, which also issues the requests to the AMO buffer. Once all the stores have drained, and the AMO buffer is in the commit stage, the request is issued to the Data cache. The buffer then receives an acknowledgement response from the cache subsystem once the request has been served. This buffer is also being flushed when its content is not being committed.

### 4.2. Introduction to the MSHR

This section will now detail two important notions: the Miss Status Holding Register (MSHR) and read/write stalls.

The MSHR is a structure used when a read-miss happens in the cache. It permits to store some information about a pending read-miss operation for later computation. The CVA6 data cache's MSHR contains the physical address of the involved miss, the size of the data, and the validity of the corresponding MSHR entry. The MSHR can only contain one entry. The ID is used and passed to other components inside the data cache and enables them to select and compute a specific cache miss.

A stall inside the cache is a specific corner-case situation where the processing of a request in the cache cannot continue because inflight requests are in contradiction or overlapping each other or another request (a request stored in the AMO buffer or the MSHR for example). In such a situation, the request's processing cannot continue and the pending requests have to be completed before being able to pursue the calculations. Otherwise, some requests might read incorrect data or overwrite entries that were not supposed to be deleted. For example, if a read-miss is pending on a given address and a write-miss is issued to the same address, there is a stall because if the pending read-miss is not completed first, then it might read the wrong data because the previous one was overwritten by the write-miss to the same address. All of this is possible in the CVA6 core because it is not fully an in-order processor. The execution stage is out-of-order and tries to optimize the usage of the execution units. Except the execution stage, which interacts a lot with the caches, the rest of the CPU is in order. Therefore, there are two possible stall situations:

    1) A write-miss collides with the MSHR (read-miss) address

    2) A read-miss cache way address overlaps with a write contained in the AMO buffer that is in flight

### 4.3. Handling misses and more with the miss unit

Now that the MSHR and stalls have been defined, it is possible to study the miss unit's role. The miss unit is a very important part of the data cache subsystem. It is responsible for the handling of data requests when there is a cache miss. It plays several key roles:

- Checking for read/write stalls
- Managing the cache lines' replacement for read operations (pseudo-random policy previously detailed)
- Flushing the cache
- Communicating with the main memory
- Communicating with the write buffer
- Writing to cache memory

The miss unit is the only component of the data cache subsystem that communicates with the CPU's main memory (located outside of the Data cache). It issues outgoing memory requests. The main memory also sends information to the miss unit. The different signals received from the main memory enable it to keep track of the pending stores, and to get responses to the issued requests: acknowledgements, validity of the requests, pass or fail status. In case of cache misses, the miss unit writes all the requests to memory that are not colliding with MSHR addresses back to the cache once the data is received from it. Additionally, the miss unit transmits the IDs (contained in the MSHR) of the cache misses back to the write buffer. The write buffer is in charge of checking the data cache memory and notifies the miss unit when a cache miss happens and passes all the required information to handle the miss correctly.

The miss unit's behavior is represented on the Figure 9. It has seven possible states: Idle, Drain, AMO, Flush, Store_Wait, Load_Wait, AMO_Wait. Idle is the "waiting state" where the miss unit is inactive and waiting for a trigger signal. In the drain state, the miss unit will wait for the write buffer and the MSHR to be cleared. The AMO state corresponds to the miss unit sending out an AMO operation request to the AMO buffer. In the flush state, the miss unit will trigger a cache flush. Store_Wait and Load_Wait correspond to a waiting state where the miss unit is idle until the corresponding request (load or store) is acknowledged by the cache memory. In the AMO_Wait state, the miss unit is stalled and waits until the pending AMO operation returns.

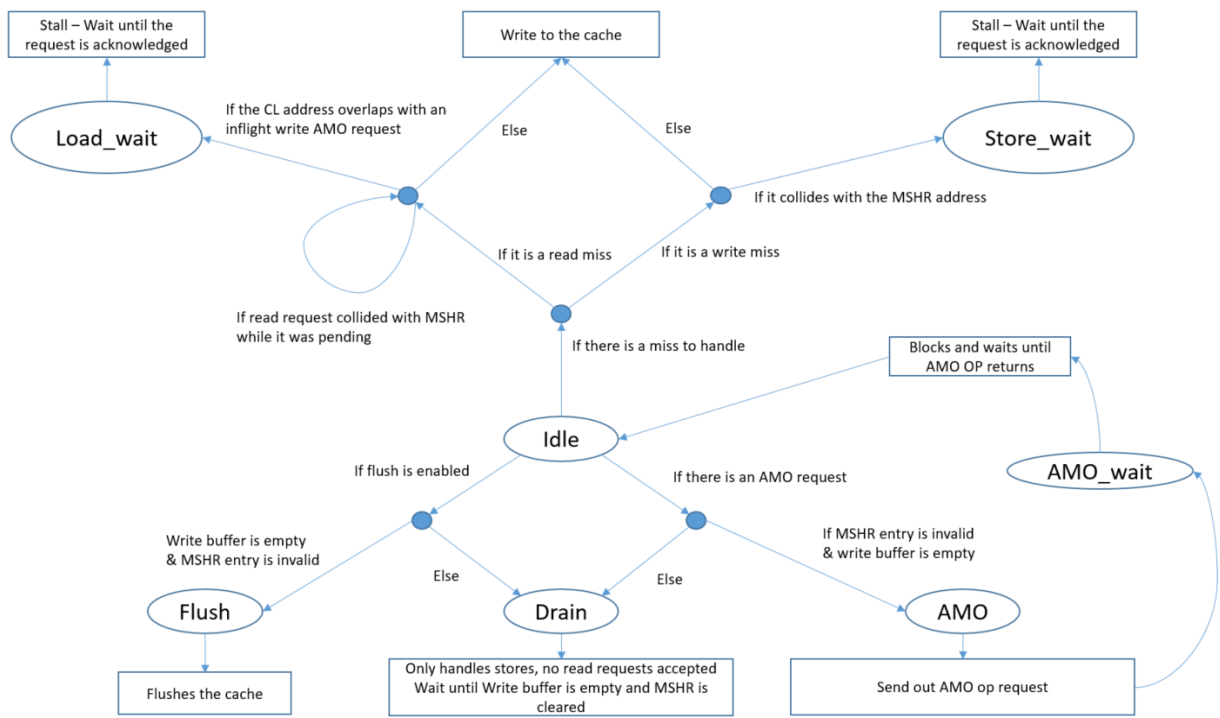

*Figure 9 - Miss unit's Final State Machine*

### 4.4. The write buffer's different usages

This buffer behaves as a fully associative cache. This also means that this cache is coalescing. This means that two adjacent free blocks of cache memory are merged. The write buffer is a central component for the Data cache as it communicates with the Cache's memory, the CPU and the Miss unit.

The write buffer is connected to the CPU through the Load and Store Unit (LSU) that is contained in the "CPU" block on the Figure 10. More specifically, the load unit, which is a component of the LSU, is directly connected to the data cache (D$) through the Write buffer as shown on the Figure 10. The role of the write buffer is to forward data to the load unit for further processing in the pipeline. The Data cache memory containing the raw data stored in the cache is not directly connected to the rest of the CPU. Therefore, the Write Buffer is in charge of transmitting the requested data from the Data cache to the rest of the CVA6's components in case of a cache hit. This means that the Write Buffer is in charge of updating the already allocated cache lines. In case of a cache miss, it is directly connected to the Miss unit and passes the request to this component for further handling. The write buffer can write either a whole cache line or a single word. It can also store several pending transactions, up to DATA CACHE_MAX_TX which can be defined accordingly to one's needs. Overall, the Write Buffer is a small cache used to store read requests issued by the CPU to the data cache and is responsible for reading entries in the Data cache memory, updating them when possible, and returning the read data back to the CVA6's execution stage of the pipeline.

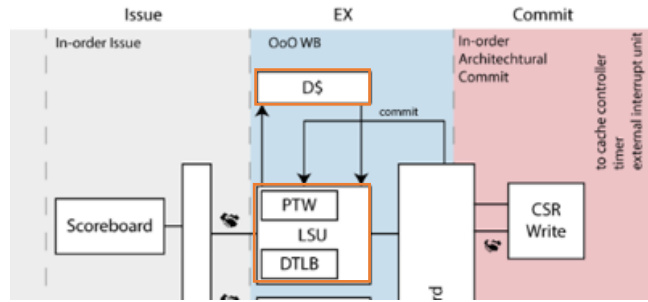
*Figure 10 - Write buffer/Load Unit interconnection in the CVA6 pipeline*

Each byte in the Write Buffer can have several states. Each state is defined by a combination of three bits representing the byte's status: valid, dirty and txblock. The "txblock" bit is set to one in case a byte is currently involved in a memory transaction. It enables to identify inflight bytes. The Table 1 shows the different possible states for a byte in the Write Buffer.

| Valid | Dirty | Txblock | Status | Conclusion |
|---|---|---|---|---|
| 0 | 0 | 0 | Invalid | The entry in the buffer needs to be freed |
| 1 | 1 | 0 | Valid and dirty | Byte is not part of an inflight transaction |
| 1 | 0 | 1 | Valid and not dirty | Byte is part of an inflight transaction |
| 1 | 1 | 1 | Valid and dirty and part of a transaction | The byte has been overwritten while inflight and needs to be retransmitted once that byte's write returns |
| 1 | 0 | 0 | Clean state | This state is never reached in the current implementation of the Write Buffer |

*Table 1 - Table representing the different states of the bytes contained in the Write Buffer*

The previously introduced states are mostly used to distinguish between bytes that have been written and not sent to the memory subsystem yet, and bytes that are part of an ongoing transaction. When considering a byte, its current status is checked, and it is then computed accordingly.

# 5. Conclusion

After learning that some micro architectural covert channels have been replicated on the CVA6 CPU such as Nils Wistoff's work on replicating Prime+Probe, we set out to study it in order to find the root cause of such attacks, and potentially some mitigations.
The current CVA6's official documentation does not contain many details about the data cache and the related components. We concluded that it would be impossible to replicate the latest micro architectural attacks or to propose defensive solutions against them without further knowledge about the CVA6 CPU. We replicated some cache-related covert channels to observe more specifically the core's behavior. This paper describes our findings while studying the CVA6 core, and especially the specificities of its Data cache. We hope that it will help understanding the fundamental issues that lead to micro architectural attacks and correct them. We also hope that our work will give more weight to open-sourced architectures, which can be more easily corrected and adapted than "closed" ones when facing such issues as the micro architectural threats.